\documentclass[aps,twocolumn,groupedaddress,showpacs]{revtex4}

\usepackage{graphicx}
\usepackage{hyperref}
\usepackage{dcolumn}
\usepackage{bm}
\usepackage{epsfig,amsbsy,bm,amssymb}

\newcommand{\isum}%
{\mathop{\hbox{$\displaystyle\sum\kern-13.2pt\int\kern1.5pt$}}}
\renewcommand{\r}{{\bm r}}

\newcommand{\p}{{\bm p}}
  \newcommand{\A}{{\bm A}}

  \newcommand{\e}{{\bm e}}

\newcommand{\ms}{\vs{-5mm}}

\newcommand{\bt}{\begin{tabular}}
\newcommand{\et}{\end{tabular}}

  \newcommand{\w}{{\omega}}
  
\newcommand{\Wcm}[2]{
$\rm {#1}\times10^{{#2}}~W/cm^2$}

\newcommand{\eref}[1] {(\ref{#1})}
\newcommand{\Eref}[1] {Eq.~(\ref{#1})}

\newcommand{\Fref}[1] {Figure \ref{#1}}
\newcommand{\Tref}[1] {Table \ref{#1}}

\newcommand{\np}{\newpage}
\renewcommand{\H}{H$_2$~}
\newcommand{\Ne}{Ne$_2$~}
  \newcommand{\Hp}{H$_2^+$~}
  
\newcommand{\br}{\begin{eqnarray*}}
\newcommand{\er}{\end{eqnarray*}}
\newcommand{\ba}{\begin{eqnarray}}
\newcommand{\ea}{\end{eqnarray}}
\newcommand{\be}{\begin{equation}}
\newcommand{\ee}{\end{equation}}

\newcommand{\vs}{\vspace*}
\newcommand{\bp}{\begin{minipage}}
\newcommand{\ep}{\end{minipage}}
  \newcommand{\nn}{\nonumber}

\begin{document}
\bibliographystyle{apsrev}

\title {Effect of the finite speed of light  in ionization
  of extended systems}

\author{I. A. Ivanov$^{1}$}
\email{igorivanov@ibs.re.kr}
\author{Anatoli S. Kheifets$^2$}
\author{Kyung Taec Kim$^{1,3}$}

\affiliation{$^1$Center for Relativistic Laser Science, Institute for
Basic Science, Gwangju 61005, Korea}
\affiliation{$^2$Research School of Physics,
  The Australian National University, Canberra ACT 2601, Australia}
\affiliation{$^3$Department of Physics and Photon Science, GIST, Gwangju 61005, Korea}

\date{\today}

\begin{abstract}

We study propagation effects due to the finite speed of light in
ionization of extended systems. We present a general quantitative
theory of these effects and show under which conditions such effects
should appear.  The finite speed of light propagation effects are
encoded in the non-dipole terms of the time-dependent Shr\"odinger
equation and display themselves in the photoelectron momentum
distribution projected on the molecular axis. Our numerical modeling
for the \Hp molecular ion and the \Ne dimer shows that the finite
light propagation time from one atomic center to another can be
accurately determined in a table top laser experiment which is much
more readily affordable than an earlier synchrotron measurement by
Grundmann {\em et al} [Science 370, 339 (2020)].

\end{abstract}

\maketitle

Every quantum system evolves on its characteristic time scale which
varies widely between molecules (femtoseconds -- $10^{-15}$~s)
\cite{PNAS}, atoms (attoseconds -- $10^{-18}$~s) \cite{kri} and
nuclei (zeptoseconds -- $10^{-21}$~s) \cite{PRL2002}.  A crossover
between these time scales is very rare in nature. It was therefore
quite unexpected to discover an ionization process in the hydrogen
molecule that evolved on a zeptosecond time scale
\cite{delay_ds}. An explanation of this phenomenon appeared to be
quite simple. While it takes tens of attoseconds for an electron to
trespass the \H molecule, the incoming light wave sweeps from one
molecular end to another orders of magnitude faster. This results in
one of the constituent hydrogen atoms getting ionized a fraction of
the attosecond sooner than its counterpart. Such a tiny ionization
delay manifests itself quite noticeably in the two-slit electron
interference that the \H molecule readily displays \cite{CF66}.
To discover a zeptosecond delay in molecular photoionization,
\citet{delay_ds} needed to deploy an extremely bright synchrotron
source of highly energetic photons \cite{Viefhaus2013}. 
We demonstrate that even a table top laser experiment is
capable of detecting a similar effect making it much more readily
affordable. 

In this Letter we present a general quantitative theory of the delay
due to the finite speed of light propagation and we show under which
conditions such effects should manifest themselves.  In our numerical
demonstrations, we consider the \Hp molecular ion and the \Ne
dimer. The \Hp molecular ion has been scrutinized since the early days of quantum
mechanics \cite{Pauling1928} and was recently used as a model for the study
of the interference  effects in photon-momentum transfer for the process 
of molecular photoionization \cite{stefi}. Photoionization studies of \Ne is a
novelty \cite{Kunitski2019}.  We subject both targets to an attosecond
laser pulse that can be readily produced in high-order harmonics
generation sources \cite{PhysRevX.7.041030,Ren2018}. The photoelectron
flux encodes the timing information about the ionization process. This
flux is reconstructed by solving the laser-driven time-dependent
Schr\"odinger equation (TDSE). The numerical results obtained for \Hp
using the TDSE can be interpreted in a transparent qualitative way by
considering a very simple heuristic tight-binding model (TBM). This
gives us a tool for understanding the time delay caused by the finite
speed of light propagation. We apply this tool to \Hp and \Ne and find
the time delay of a fraction of attosecond that depends sensitively on
the orientation of the molecular axis relative to the propagation and
polarization directions.

Our approach is based on the numerical solution of the
three-dimensional TDSE 
\begin{equation}
i \partial \Psi(\r,t) / \partial t=
\left[\hat H_{\rm mol} + \hat H_{\rm int}(t)\right] \Psi(\r,t) \ ,
\label{tdse}
\end{equation}
where $\hat H_{\rm mol}$ is the field-free one-electron Hamiltonian and
$\hat H_{\rm int}(t)$ describes the field-target interaction. We
consider a traveling wave propagating in the positive $x$-direction
and polarized in the $z$-direction, described by the vector potential
$ \A(t-x/c)$ with $c=137.036$ being the speed of light in atomic
unites (a.u.).

We apply the procedure previously used in \cite{ndim} to study
non-dipole effects in atomic photoionization.  The leading order
relativistic corrections to $\hat H_{\rm int}(t)$ come from the linear
term of the expansion of $ \A(t-x/c)$ in powers of $c^{-1}$.  We plug
this expansion into the standard minimal coupling Hamiltonian
\cite{Sobelman72}
$\hat H_{\rm min}=  (\hat{\p} + \e_z
A(x,t))^2/2$ 
and keep the terms linear in $c^{-1}$: 
\be \hat H_{\rm int}(\r,t) = \hat p_z A(t)+ {\hat p_z x
  E(t)\over c}+ {A(t)E(t)x\over c} + {A^2(t)\over 2}\ .
\label{h1} 
\ee
In the above expression, $A(t)$ no longer depends on the coordinates
and $ E(t)=-\partial A(t)/ \partial t$ is the electric field of the
pulse.  The last term on the r.h.s. of \Eref{h1} is a function of time
only and can, therefore, be removed by a unitary transformation.  The
interaction Hamiltonian \eref{h1} can be related by a gauge
transformation to the Kramers-Henneberger Hamiltonian used in
\cite{forre}.

An additional source of relativistic corrections in $\hat H_{\rm int}$
is the interaction of the magnetic field of the pulse with the
electron spin.  These corrections cannot be
obtained by a simple generalization of the minimal coupling
Hamiltonian. To obtain them, one needs to consider systematically the 
transition of the Dirac equation to the non-relativistic limit
\cite{LL4}. The inclusion of the electron spin, however,
is not necessary for the present study where we keep only the $c^{-1}$
terms as was done, for instance, in \cite{ndi1,ndim}.  The Breit-Pauli
relativistic corrections to $\hat H_{\rm mol}$, such as the effects
of the relativistic kinematics, spin-orbit interaction and the
so-called Darwin term, are all of the order of $c^{-2}$
\cite{Sobelman72} and can also be safely omitted.

The target interacts with a short laser pulse
\be A(t)= -E_0 \omega^{-1}
\sin^2(\pi t/ T_1)\sin{\omega t} \ ,
\label{pulse}
\ee
of the total duration $T_1=4T$ with $T=2\pi/\w$ and the base laser
frequency $\omega=4.04$~a.u. (the photon energy of $110$~eV). The peak
electric field strength $E_0=0.1$~au corresponds to the intensity of
\Wcm{3.5}{14}

To solve the TDSE \eref{tdse} we follow the procedure that we employed
previously in the non-relativistic case \cite{cuspm,circ6,circ7}.  A
solution of the TDSE is sought in the spherical coordinates
as an expansion over the spherical harmonics  basis 
\be
\Psi({\bm r},t)=
\sum\limits_{l,m}^{l_{\rm max}}
f_{lm}(r,t) Y_{lm}(\hat r)  \ ,
\label{basis}
\ee
The radial grid is discretized with the step $\delta r=0.05$ a.u. in a
box of the size $R_{\rm max}=400$~au which, together with
$l_{\max}=10$, ensured a numerical convergence .
We characterize the final photoelectron state by its asymptotic
momentum $\p$. Ionization amplitudes $a_{\p}$ are obtained by
projecting the TDSE solution after the end of the pulse on the ingoing
scattering states $\phi^{-}_{\p}$ \cite{LL3}.

For a weak electromagnetic field that we employ, we can also obtain an
analytical expression for the ionization amplitude by using the
perturbation theory (PT) formula and treating the operator \eref{h1} as a
perturbation:
\be
a_{\p}= -i\int_{-\infty}^{+\infty} \langle \phi^{-}_{\p}|\hat H_{\rm int}(\tau)|\phi_0\rangle 
e^{i(E_{\p}-\varepsilon_0)\tau}\ d\tau \ .
\label{a1}
\ee
Here $\phi_0$ and $\phi^-_{\p}$ are the initial and final molecular
states with the corresponding energies $\varepsilon_0$ and $E_{\p}$.

We split the ionization amplitude
\eref{a1} 
$
a_{\p}= a^{(0)}_{\p} + a^{(1)}_{\p} 
$
into the nonrelativistic part $a^{(0)}_{\p}$ and the first order
relativistic correction $a^{(1)}_{\p}$. By introducing the Fourier
transforms
$A(t) = (2\pi)^{-1} \int a(\Omega) 
e^{-i t\Omega}\ d\Omega
$
and 
$A^2(t) = (2\pi)^{-1}
\int b(\Omega) e^{-i t\Omega}\ d\Omega$ \,
we obtain for these amplitudes:
\ba
a^{(0)}_{\p}&=& 
-i a(\Omega) \langle \phi^-_{\p}|\hat p_z|\phi_0 \rangle\ 
\ \ , \ \ 
\Omega= E_{\p}- \varepsilon_0
\label{a3}
\\
a^{(1)}_{\p}&=& \Omega c^{-1}
\left(a(\Omega)\langle \phi^-_{\p}|
\hat p_z x|\phi_0 \rangle\ + b(\Omega) 
\langle \phi^-_{\p}|x |\phi_0 \rangle/2 \right)  \nonumber \\ 
&\approx & \Omega c^{-1}a(\Omega) \langle \phi^-_{\p}|
\hat p_z x|\phi_0 \rangle \ .
\label{a4}
\ea
The second term on the right hand side of \Eref{a4} can be neglected
because the ratio of the second term and the first term in this
expression is approximately $b(\Omega)/(a(\Omega) p)$, where $p$ is
the typical value of the momentum of the ionized electron. For the
pulse parameters that we consider, this value can be estimated as
$E_0/(\omega p)\approx 0.01$.

It can be seen from \Eref{a4} that the relativistic correction
$a^{(1)}_{\p}$  vanishes in certain cases. In particular, it is zero
for an axially symmetric system with the field-free Hamiltonian that is
invariant under rotations about the $z$-axis.  Indeed, for such
systems the scattering state $\phi^-_{\p}(\r)$ is a function of the
arguments $z, p_z, \rho, p_{\rho}$ and ${\bm\rho}\cdot\p_{\rho}$,
where $z$, $p_z$ and ${\bm\rho}$, $\p_{\rho}$ are the components of
the $\r$ and $\p$ vectors parallel and perpendicular, respectively, to
the $z$- axis of the coordinate system that we employ. 
For an axially symmetric system the ground state wave function 
$\phi_0$ in \Eref{a4} (which we assume to be non-degenerate) is
invariant under rotations about the $z$-axis. We obtain then from
\Eref{a4}: $\displaystyle
a^{(1)}(p_x,p_y,p_z)=-a^{(1)}(-p_x,p_y,p_z)$. Therefore the amplitude
$a^{(1)}_{\p}$ is zero in the plane $p_x=0$ which encompasses the most
important region of the momenta near the maximum of the momentum
distribution.

The effect of the propagation correction \eref{a4} can be most easily
illustrated within the tight-binding model in which the ground state
of \Hp and \Ne is represented by a Heitler-London wave function
$$ 
\phi_0(\r)= \large[\phi\large(\r-{\bm{R}/ 2}\large) +
\phi\large(\r+{\bm{R}/ 2}\large) \large]/\sqrt{2} 
\ .
$$
In the TBM, the overlap of the two terms is small and $\phi(\r)$ is
represented by a spherically symmetric atomic-like state.  Under these
conditions, and by employing the Born approximation for the
scattering state $\phi^-_{\p}$, which is justified for the relatively
high electron energies considered here, we obtain:
\ba
a^{(0)}_{\p}&=& a_d(\p) \sqrt{2}\cos{\left({\p\cdot {\bm R}/ 2}\right)} \ ,
\label{a5}
\\
a_d(\p)&=& -i a(\Omega) \langle \p|\hat p_z|\phi \rangle\
\label{ad}
\\
a^{(1)}_{\p}&=& -\Omega c^{-1} (\partial/ \partial p_x) a^{(0)}_{\p} \ .
\label{a6}
\ea
The non-relativistic term $a^{(0)}_{\p}$ is represented by the
single-center dipole amplitude \eref{ad} modulated by the two-center
interference factor \cite{CF66}. Adding the relativistic term
leads to
\be
a_{\p}= a^{(0)}_{\p} + a^{(1)}_{\p} \approx 
a_d(\p) \sqrt{2} 
\cos{\left({\p\cdot {\bm R}/ 2} +\delta\right)} \ ,
\label{a7}
\ee
where $\delta =- {\bm \kappa}\cdot{\bm R}/2$ and ${\bm \kappa}=\Omega
c^{-1}\e_x$ is the photon momentum. It is this phase factor $\delta$
that is responsible for the modified interference pattern observed and
decoded in \cite{delay_ds}.

\begin{widetext}
\begin{figure*}
\resizebox{160mm}{!}{\epsffile{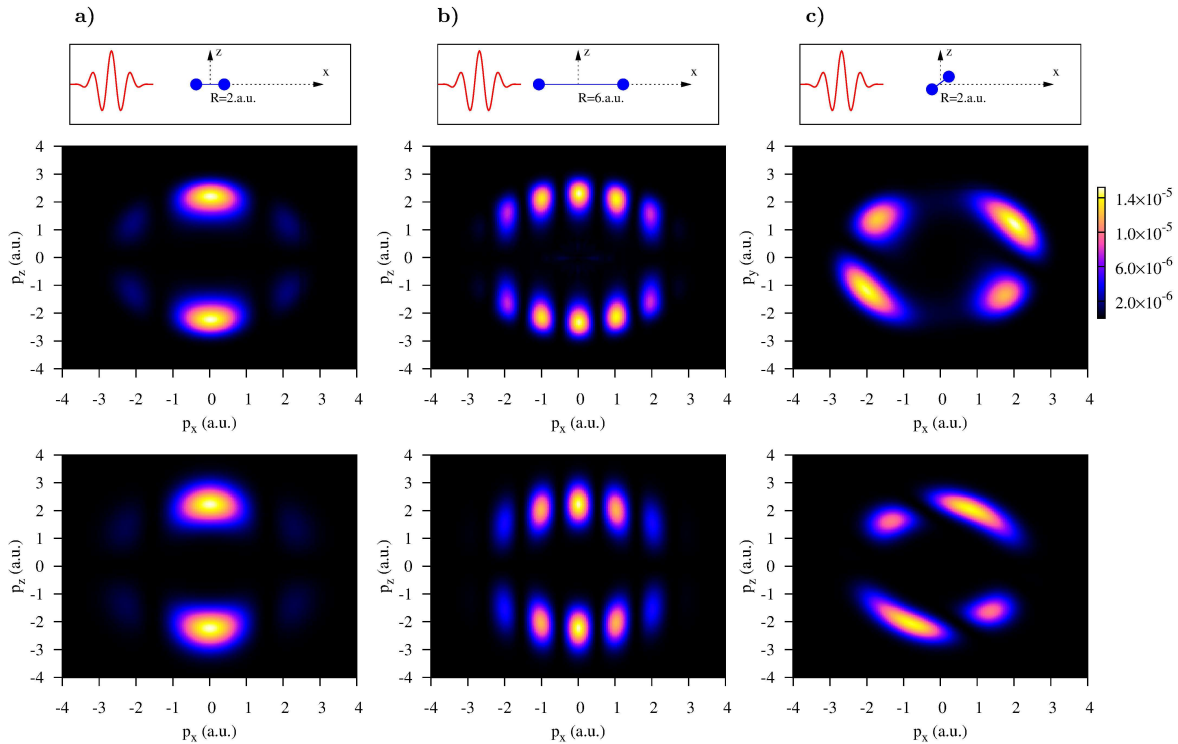}}

\caption{(Color online) The photoelectron momentum distributions
  projected on the $(p_x,p_y)$-plane for different orientations and
  inter-nuclear distances of \Hp illustrated in the top row of
  panels. The middle row of panels displays the numerical TDSE results
  while the bottom row exhibits the corresponding TBM results
  obtained using \Eref{a5}.}
\label{fig1}
\vs{-5mm}
\end{figure*}
\end{widetext}

Appearance of the additional phase factor $\delta$ in \Eref{a7} is due
to an extra propagation time of the light wave from one end of the
molecule to the other. This interpretation becomes yet more
transparent if we use coordinate representation for the part of the 
wave-function describing the ionized wave packet:

\be
\Psi_{\rm ion}(\r,t)= \int a_{\p} \phi^{-}_{\p}(\r) e^{-iE_{\p}t}\ d\p \ ,
\label{a8}
\ee

where $\phi^{-}_{\p}(\r)$ are the molecular scattering states. To
evaluate this integral in the limit $t\to\infty$, we rely on the
saddle-point method (SPM) that is commonly used in description of
ionization \cite{mined2,delay1}  or 
scattering \cite{gw64} processes.
By writing $a_d(\p)=|a_d(\p)|e^{i\eta(\p)}$ in \Eref{a7} and
employing the SPM, we obtain from \Eref{a8}:

\begin{eqnarray}
\nn
\lim\limits_{t\to\infty} \Psi_{\rm ion}(\r,t)&=&e^{i\p_{\rm m}\cdot \r} 
\Big\{ G\left[\r+{{\bm R}/2}-\p_{\rm m}(t-\tau-\tau_1)\right] 
\\&+& G\left[\r-{{\bm R}/ 2}-\p_{\rm m}(t-\tau+\tau_1)\right]\Big\} \ . 
\label{a9}
\end{eqnarray}

Here $\p_{\rm m}$ is the most probable photoelectron momentum,
$\displaystyle \tau= {\partial \eta/ \partial E}$ is the usual Wigner
photoemission time-delay \cite{delay2} and $ \tau_1= {\partial
\delta/ \partial E}= -R_x/(2c)$ with $\delta =- {\bm \kappa}\cdot{\bm R}/2$
from \Eref{a7}. 

The $\tau_1$ term represents an additional time that
it takes for the light pulse to cover the distance $R_x= {\bm R}\cdot
\e_x$. 

The wave packet $G(\r)$ in \Eref{a9} is a Fourier transform of
the dipole amplitude $a_d(\p)$ in \Eref{a7}.  Importantly, $G(\r)$ has
a strong peak near the origin. Therefore the two terms in \Eref{a9}
describe two wave packets emitted from the two atomic centers $\r=-\bm
R/2$ and $\r=\bm R/2$ at the times $\tau+ \tau_1$ and $\tau-\tau_1$,
respectively.  For transparency of derivation, we omitted in \Eref{a9}
the Coulomb terms which would only add slowly varying (logarithmic)
corrections in the arguments of $G(\r)$ \cite{mined2,delay1}. These
additional logarithmic terms are the same for the two wave packets and
they would therefore cancel in the time delay difference between these
wave packets.

We note that the right-hand side of \Eref{ad} can be represented as
a product of two factors,  the factor $p_z$ responsible for the angular dependence
of the amplitude and the Gaussian factor
$\exp{\left\{-a(p-p_0)^2\right\}}$ representing the energy
conservation $p_0^2/2= \varepsilon_0+ \omega$. Such a Gaussian
representation of the ionized wave packets emitted in the
single-center problems is often used in studying temporal dynamics of
atomic ionization \cite{delay1,mined2}. The parameter $a$ determines the
width of the wave packet and depends on the pulse parameters. 
One can derive an analytical expression for its value but
we will not need it in the following.  By employing the Gaussian
ansatz for $a_d(\p)$ and \Eref{a7} we finally obtain
\be a_{\p} = A \exp{\left\{-a(p-p_0)^2\right\}} p_z
\cos{\left({\p\cdot {\bm R}/ 2} +\delta\right)} \ .
\label{f1}
\ee

We use \Eref{f1} to evaluate the photoelectron emission pattern and to
compare it with the fully {\em ab initio} TDSE calculations for
various orientations and different inter-nuclear distances of \Hp.
This comparison is presented in \Fref{fig1}. The top row of panels
illustrates the geometry of the ionization process. It is assumed that
the molecular axis is confined to the $xz$-plane where the propagation
and polarization vectors of the pulse belong, making an angle $\theta$
with the propagation direction. In (a-b), $\theta=0$ while in (c)
$\theta=\pi/4$. The photoelectron momentum distribution is projected on
the $xz$ plane and computed as $P(p_x,p_z)=\int |a_{\p}|^2\ dp_y$ with
the amplitude $a_{\p}$ obtained by projecting the TDSE solution on the
set of the scattering states of \Hp. The number and location of the
bright spots in \Fref{fig1} reflect a simple two-center interference
pattern governed by the cosine term in \Eref{f1}.

\begin{figure}[!]
\resizebox{80mm}{!}{\epsffile{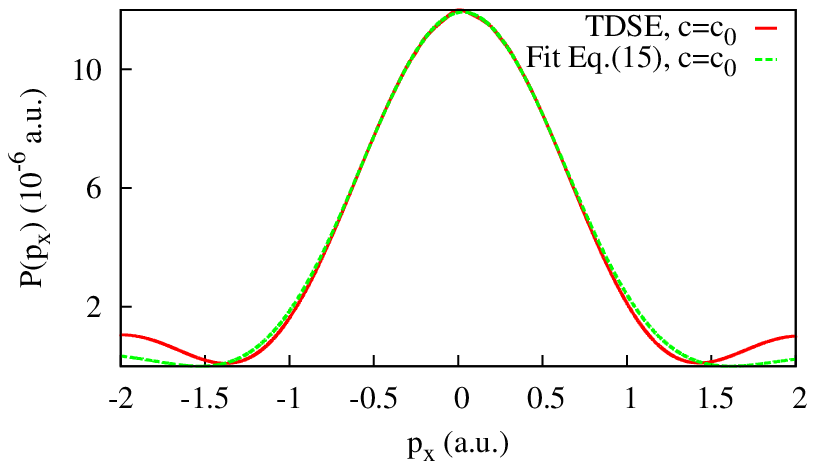}} 
\resizebox{80mm}{!}{\epsffile{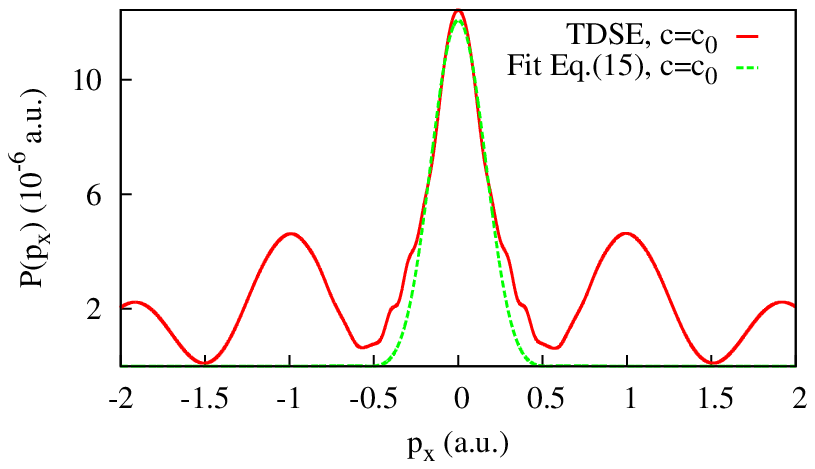}} 
\caption{(Color online) The photoelectron momentum distribution
  projected on the molecular axis for geometries illustrated in
  \Fref{fig1}a) (top) and 1b) (bottom). The fit based on \Eref{fit}
  and TDSE calculations are plotted with the solid and dotted lines,
  respectively. } \ms
\label{fig2}
\end{figure}

Comparison of the TDSE calculations (the middle row of panels in
\Fref{fig1}) and results based on \Eref{f1} (the bottom row of panels)
shows that the TBM reproduces the spectra very well for the geometries
that we consider. We will, therefore, analyze and interpret our TDSE results using
this transparent model that is equally applicable to both \Hp and \Ne.  
We will focus on the photoelectron momentum distribution
projected on the $(xz)$ plane and integrated over the momentum
component $p_{\perp}$ which is perpendicular to the molecular
axis. Such a momentum distribution is a function of the 
momentum component $p_{||}$ which is parallel to the molecular
axis. By employing the Gaussian ansatz \eref{f1} we obtain: 
\begin{eqnarray}
\label{fit}
P(p_{||})&=& \int P(\p)\ dp_y dp_{\perp} \\ & \approx & B
\exp{\left\{-C p_{||}^2\right\}} \cos^2{\left({p_{||}R\over 2} +
  \delta\right)} \ .  \nonumber
\end{eqnarray}
We use the analytical expression \eref{fit} with adjustable
parameters $B$, $C$ and $\delta$ to fit $P(p_\|)$ obtained from the
numerical TDSE calculations.  The accuracy of the fitting procedure is
illustrated in \Fref{fig2}a) and b) were we display $P(p_\|)$ for the
two geometries illustrated in \Fref{fig1}a) and b), respectively. In
both cases, $\theta=0$ and $P(p_\|)$ is simply the projection of the
2D momentum distribution on the horizontal axis.
%
%
As is seen in \Fref{fig2}, the analytical fit with
\Eref{fit} represents the TDSE calculation with the corresponding 
$R$ value  quite accurately. This allows us to
extract the phase shift $\delta$ accumulated due to the finite speed
of light propagation.

\begin{figure}[!]
	\begin{tabular} {l}
\resizebox{80mm}{!}{\epsffile{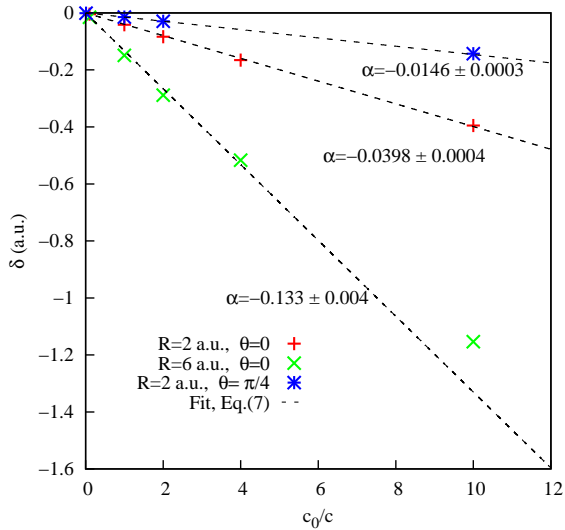}} 
\end{tabular}
\ms
\caption{(Color online) Parameter $\delta$ in \Eref{fit} for 
different geometries and internuclear distances.}
\label{fig3}
\ms
\end{figure}

In \Fref{fig3} we display thus extracted phase shifts $\delta$ for the
three molecular orientations illustrated in \Fref{fig1}. To enhance the
relativistic effects, we artificially decrease the speed of light from its
physical value $c=c_0=137.036$~a.u.  down to  $c=c_0/10$. 
According to TBM, the phase shift should scale
linearly as $\delta= \alpha c_0/c$ with the slope $\alpha=
-R_x\kappa/2$.  The predicted linear scaling is confirmed very
accurately by the numerical values shown in \Fref{fig3} with only a
small error margin. 
%
%
The time delay values corresponding to the physical speed of
light $c_0$ are shown in \Tref{tab1} in comparison with the
estimate $\Delta t= 2R_x/c_0$ provided by the TBM. Agreement of the results
can be deemed quite satisfactory given the relative simplicity of TBM.
More importantly, the linear dependence of the TDSE results on the
parameter $c_0/c$ clearly demonstrates existence of the finite speed
of light effect in ionization of diatomic molecules.

\begin{table}
\caption{\label{table1} Propagation delay for different geometries and internuclear distances.}
\begin{ruledtabular}
\begin{tabular}{lccr}
R (a.u.) & $\theta$  & Fit (as)  & $R_x/c$ (as) \\
\hline
 2    &  0  & 0.46 & 0.35  \\
 2    &  $\pi/4$  & 0.17 & 0.24 \\
6	  &  0 &  1.56 & 1.05 \\
\end{tabular}
\end{ruledtabular}
\label{tab1}
\end{table}

In conclusion, our work was motivated by the synchrotron based
experiment by \citet{delay_ds} who discovered a zeptosecond time delay
in photoionization of the \H molecule. We aimed to demonstrate that a
similar delay, which is caused by the finite speed of light
propagation from one constituent atom to another, can be detected in a
much more accessible table top laser settings. In doing so, we
developed a general theory of the finite speed of light propagation
effects in ionization of extended systems.  As a simple case study, we
considered the \Hp molecular ion interacting with a short laser
pulse. This target affords a very accurate numerical treatment within
the time-dependent Schr\"odinger equation. At the same time, an
heuristic tight-binding model employing the Heitler-London molecular
ground state produces very similar results. TBM can be easily adopted
to the \Ne dimer by a simple increase of the inter-atomic distance
to $R\simeq6$~a.u. Notably, the corresponding interference pattern
displayed in \Fref{fig1}b) is very similar to that obtained in the
experiment \cite{Kunitski2019} and the  earlier SPM modeling
\cite{PhysRevA.99.063428}.

Our simulations demonstrate that the speed of light delay in
ionization of diatomic molecular targets can be decoded from the
photoelectron momentum distribution projected on the molecular
axis. This method has a clear advantage over the earlier synchrotron
measurement based on decoding the 2D interference pattern
\cite{delay_ds}. Indeed, the bright interference spots have a finite
angular width. To detect their shift due to the finite speed of light
requires a sufficiently large photon momentum that should not be
vanishingly small in comparison with the photoelectron momentum. This
in turn requires very high photon energy (800~eV in
\cite{delay_ds}). Fitting of a one-dimensional momentum distribution
produces significantly reduced error bars. Thus smaller values of the
photon momentum and energy can be used (100~eV in our simulations). In
addition, by projecting the momentum distribution on the molecular
axis, we increase the count rate and improve statistics of the
measurement. Hence the photon flux can be significantly reduced. This
reduction of both the photon flux and energy makes the proposed method
much more readily accessible in desk-top conventional laser
settings. This we hope will stimulate further speed of light delay
determinations in diatomic molecules and other extended systems.

\np



\begin{thebibliography}{22}
\expandafter\ifx\csname natexlab\endcsname\relax\def\natexlab#1{#1}\fi
\expandafter\ifx\csname bibnamefont\endcsname\relax
  \def\bibnamefont#1{#1}\fi
\expandafter\ifx\csname bibfnamefont\endcsname\relax
  \def\bibfnamefont#1{#1}\fi
\expandafter\ifx\csname citenamefont\endcsname\relax
  \def\citenamefont#1{#1}\fi
\expandafter\ifx\csname url\endcsname\relax
  \def\url#1{\texttt{#1}}\fi
\expandafter\ifx\csname urlprefix\endcsname\relax\def\urlprefix{URL }\fi
\providecommand{\bibinfo}[2]{#2}
\providecommand{\eprint}[2][]{\url{#2}}

\bibitem[{\citenamefont{Remacle and Levine}(2006)}]{PNAS}
\bibinfo{author}{\bibfnamefont{F.}~\bibnamefont{Remacle}} \bibnamefont{and}
  \bibinfo{author}{\bibfnamefont{R.~D.} \bibnamefont{Levine}},
  \emph{\bibinfo{title}{An electronic time scale in chemistry}},
  \bibinfo{journal}{Proc. Nat. Acad. Sci.}
  \textbf{\bibinfo{volume}{103}}(\bibinfo{number}{18}), \bibinfo{pages}{6793}
  (\bibinfo{year}{2006}).

\bibitem[{\citenamefont{Krausz and Ivanov}(2009)}]{kri}
\bibinfo{author}{\bibfnamefont{F.}~\bibnamefont{Krausz}} \bibnamefont{and}
  \bibinfo{author}{\bibfnamefont{M.}~\bibnamefont{Ivanov}},
  \emph{\bibinfo{title}{Attosecond physics}}, \bibinfo{journal}{Rev. Mod.
  Phys.} \textbf{\bibinfo{volume}{81}}, \bibinfo{pages}{163}
  (\bibinfo{year}{2009}).

\bibitem[{\citenamefont{Kaplan and Shkolnikov}(2002)}]{PRL2002}
\bibinfo{author}{\bibfnamefont{A.~E.} \bibnamefont{Kaplan}} \bibnamefont{and}
  \bibinfo{author}{\bibfnamefont{P.~L.} \bibnamefont{Shkolnikov}},
  \emph{\bibinfo{title}{Lasetron: A proposed source of powerful
  nuclear-time-scale electromagnetic bursts}}, \bibinfo{journal}{Phys. Rev.
  Lett.} \textbf{\bibinfo{volume}{88}}, \bibinfo{pages}{074801}
  (\bibinfo{year}{2002}).

\bibitem[{\citenamefont{Grundmann et~al.}(2020)\citenamefont{Grundmann,
  Trabert, Fehre, Strenger, Pier, Kaiser, Kircher, Weller, Eckart, Schmidt
  et~al.}}]{delay_ds}
\bibinfo{author}{\bibfnamefont{S.}~\bibnamefont{Grundmann}},
  \bibinfo{author}{\bibfnamefont{D.}~\bibnamefont{Trabert}},
  \bibinfo{author}{\bibfnamefont{K.}~\bibnamefont{Fehre}},
  \bibinfo{author}{\bibfnamefont{N.}~\bibnamefont{Strenger}},
  \bibinfo{author}{\bibfnamefont{A.}~\bibnamefont{Pier}},
  \bibinfo{author}{\bibfnamefont{L.}~\bibnamefont{Kaiser}},
  \bibinfo{author}{\bibfnamefont{M.}~\bibnamefont{Kircher}},
  \bibinfo{author}{\bibfnamefont{M.}~\bibnamefont{Weller}},
  \bibinfo{author}{\bibfnamefont{S.}~\bibnamefont{Eckart}},
  \bibinfo{author}{\bibfnamefont{L.~P.~H.} \bibnamefont{Schmidt}},
  \bibnamefont{et~al.}, \emph{\bibinfo{title}{Zeptosecond birth time delay in
  molecular photoionization}}, \bibinfo{journal}{Science}
  \textbf{\bibinfo{volume}{370}}, \bibinfo{pages}{339} (\bibinfo{year}{2020}).

\bibitem[{\citenamefont{Cohen and Fano}(1966)}]{CF66}
\bibinfo{author}{\bibfnamefont{H.~D.} \bibnamefont{Cohen}} \bibnamefont{and}
  \bibinfo{author}{\bibfnamefont{U.}~\bibnamefont{Fano}},
  \emph{\bibinfo{title}{Interference in the photo-ionization of molecules}},
  \bibinfo{journal}{Phys.~Rep.}
  \textbf{\bibinfo{volume}{150}}(\bibinfo{number}{1}), \bibinfo{pages}{30}
  (\bibinfo{year}{1966}).

\bibitem[{\citenamefont{Viefhaus et~al.}(2013)\citenamefont{Viefhaus, Scholz,
  Deinert, Glaser, Ilchen, Seltmann, Walter, and Siewert}}]{Viefhaus2013}
\bibinfo{author}{\bibfnamefont{J.}~\bibnamefont{Viefhaus}},
  \bibinfo{author}{\bibfnamefont{F.}~\bibnamefont{Scholz}},
  \bibinfo{author}{\bibfnamefont{S.}~\bibnamefont{Deinert}},
  \bibinfo{author}{\bibfnamefont{L.}~\bibnamefont{Glaser}},
  \bibinfo{author}{\bibfnamefont{M.}~\bibnamefont{Ilchen}},
  \bibinfo{author}{\bibfnamefont{J.}~\bibnamefont{Seltmann}},
  \bibinfo{author}{\bibfnamefont{P.}~\bibnamefont{Walter}}, \bibnamefont{and}
  \bibinfo{author}{\bibfnamefont{F.}~\bibnamefont{Siewert}},
  \emph{\bibinfo{title}{{The Variable Polarization XUV Beamline P04 at PETRA
  III : Optics, mechanics and their performance}}}, \bibinfo{journal}{Nucl.
  Instrum. Methods Phys. Res. A} \textbf{\bibinfo{volume}{710}},
  \bibinfo{pages}{151} (\bibinfo{year}{2013}).

\bibitem[{\citenamefont{Pauling}(1928)}]{Pauling1928}
\bibinfo{author}{\bibfnamefont{L.}~\bibnamefont{Pauling}},
  \emph{\bibinfo{title}{The application of the quantum mechanics to the
  structure of the hydrogen molecule and hydrogen molecule-ion and to related
  problems.}}, \bibinfo{journal}{Chemical Reviews}
  \textbf{\bibinfo{volume}{5}}(\bibinfo{number}{2}), \bibinfo{pages}{173}
  (\bibinfo{year}{1928}).
	
	\bibitem[{\citenamefont{Chelkowski and Bandrauk}(2018)}]{stefi}
\bibinfo{author}{\bibfnamefont{S.} \bibnamefont{Chelkowski}} \bibnamefont{and}
  \bibinfo{author}{\bibfnamefont{A.~D.} \bibnamefont{Bandrauk}},
  \emph{\bibinfo{title}{Photon-momentum transfer in molecular photoionization}}, 
	\bibinfo{journal}{Phys. Rev. A}
  \textbf{\bibinfo{volume}{97}}, \bibinfo{pages}{053401}
  (\bibinfo{year}{2018}).
	

\bibitem[{\citenamefont{Kunitski et~al.}(2019)\citenamefont{Kunitski, Eicke,
  Huber, {K\"ohler}, Zeller, Voigtsberger, Schlott, Henrichs, Sann, Trinter
  et~al.}}]{Kunitski2019}
\bibinfo{author}{\bibfnamefont{M.}~\bibnamefont{Kunitski}},
  \bibinfo{author}{\bibfnamefont{N.}~\bibnamefont{Eicke}},
  \bibinfo{author}{\bibfnamefont{P.}~\bibnamefont{Huber}},
  \bibinfo{author}{\bibfnamefont{J.}~\bibnamefont{{K\"ohler}}},
  \bibinfo{author}{\bibfnamefont{S.}~\bibnamefont{Zeller}},
  \bibinfo{author}{\bibfnamefont{J.}~\bibnamefont{Voigtsberger}},
  \bibinfo{author}{\bibfnamefont{N.}~\bibnamefont{Schlott}},
  \bibinfo{author}{\bibfnamefont{K.}~\bibnamefont{Henrichs}},
  \bibinfo{author}{\bibfnamefont{H.}~\bibnamefont{Sann}},
  \bibinfo{author}{\bibfnamefont{F.}~\bibnamefont{Trinter}},
  \bibnamefont{et~al.}, \emph{\bibinfo{title}{Double-slit photoelectron
  interference in strong-field ionization of the neon dimer}},
  \bibinfo{journal}{Nature Communications} \textbf{\bibinfo{volume}{10}},
  \bibinfo{pages}{1} (\bibinfo{year}{2019}).

\bibitem[{\citenamefont{Cousin et~al.}(2017)\citenamefont{Cousin, Di~Palo,
  Buades, Teichmann, Reduzzi, Devetta, Kheifets, Sansone, and
  Biegert}}]{PhysRevX.7.041030}
\bibinfo{author}{\bibfnamefont{S.~L.} \bibnamefont{Cousin}},
  \bibinfo{author}{\bibfnamefont{N.}~\bibnamefont{Di~Palo}},
  \bibinfo{author}{\bibfnamefont{B.}~\bibnamefont{Buades}},
  \bibinfo{author}{\bibfnamefont{S.~M.} \bibnamefont{Teichmann}},
  \bibinfo{author}{\bibfnamefont{M.}~\bibnamefont{Reduzzi}},
  \bibinfo{author}{\bibfnamefont{M.}~\bibnamefont{Devetta}},
  \bibinfo{author}{\bibfnamefont{A.}~\bibnamefont{Kheifets}},
  \bibinfo{author}{\bibfnamefont{G.}~\bibnamefont{Sansone}}, \bibnamefont{and}
  \bibinfo{author}{\bibfnamefont{J.}~\bibnamefont{Biegert}},
  \emph{\bibinfo{title}{Attosecond streaking in the water window: A new regime
  of attosecond pulse characterization}}, \bibinfo{journal}{Phys. Rev. X}
  \textbf{\bibinfo{volume}{7}}, \bibinfo{pages}{041030} (\bibinfo{year}{2017}).

\bibitem[{\citenamefont{Ren et~al.}(2018)\citenamefont{Ren, Li, Yin, Zhao,
  Chew, Wang, Hu, Cheng, Cunningham, Wu et~al.}}]{Ren2018}
\bibinfo{author}{\bibfnamefont{X.}~\bibnamefont{Ren}},
  \bibinfo{author}{\bibfnamefont{J.}~\bibnamefont{Li}},
  \bibinfo{author}{\bibfnamefont{Y.}~\bibnamefont{Yin}},
  \bibinfo{author}{\bibfnamefont{K.}~\bibnamefont{Zhao}},
  \bibinfo{author}{\bibfnamefont{A.}~\bibnamefont{Chew}},
  \bibinfo{author}{\bibfnamefont{Y.}~\bibnamefont{Wang}},
  \bibinfo{author}{\bibfnamefont{S.}~\bibnamefont{Hu}},
  \bibinfo{author}{\bibfnamefont{Y.}~\bibnamefont{Cheng}},
  \bibinfo{author}{\bibfnamefont{E.}~\bibnamefont{Cunningham}},
  \bibinfo{author}{\bibfnamefont{Y.}~\bibnamefont{Wu}}, \bibnamefont{et~al.},
  \emph{\bibinfo{title}{Attosecond light sources in the water window}},
  \bibinfo{journal}{Journal of Optics}
  \textbf{\bibinfo{volume}{20}}(\bibinfo{number}{2}), \bibinfo{pages}{023001}
  (\bibinfo{year}{2018}).

\bibitem[{\citenamefont{I.A.Ivanov et~al.}(2016)\citenamefont{I.A.Ivanov,
  J.Dubau, and Kim}}]{ndim}
\bibinfo{author}{\bibnamefont{I.A.Ivanov}},
  \bibinfo{author}{\bibnamefont{J.Dubau}}, \bibnamefont{and}
  \bibinfo{author}{\bibfnamefont{K.~T.} \bibnamefont{Kim}},
  \emph{\bibinfo{title}{Nondipole effects in strong-field ionization}},
  \bibinfo{journal}{Phys.~Rev.~A} \textbf{\bibinfo{volume}{94}},
  \bibinfo{pages}{033405} (\bibinfo{year}{2016}).

\bibitem[{\citenamefont{Sobelman}(1972)}]{Sobelman72}
\bibinfo{author}{\bibfnamefont{I.~I.} \bibnamefont{Sobelman}},
  \emph{\bibinfo{title}{Introduction to the Theory of Atomic Spectra}}
  (\bibinfo{publisher}{Pergamon Press}, \bibinfo{address}{New York},
  \bibinfo{year}{1972}).

\bibitem[{\citenamefont{F{\o}rre}(2006)}]{forre}
\bibinfo{author}{\bibfnamefont{M.}~\bibnamefont{F{\o}rre}},
  \bibinfo{journal}{Phys.~Rev.~A} \textbf{\bibinfo{volume}{74}},
  \bibinfo{pages}{065401} (\bibinfo{year}{2006}).

\bibitem[{\citenamefont{Lifshitz and Berestetskii}(1982)}]{LL4}
\bibinfo{author}{\bibfnamefont{E.~M.} \bibnamefont{Lifshitz}} \bibnamefont{and}
  \bibinfo{author}{\bibfnamefont{V.~B.} \bibnamefont{Berestetskii}},
  \emph{\bibinfo{title}{Quantum Electrodynamics}} (\bibinfo{publisher}{Pergamon
  Press}, \bibinfo{year}{1982}).

\bibitem[{\citenamefont{Chelkowski et~al.}(2015)\citenamefont{Chelkowski,
  Bandrauk, and Corkum}}]{ndi1}
\bibinfo{author}{\bibfnamefont{S.}~\bibnamefont{Chelkowski}},
  \bibinfo{author}{\bibfnamefont{A.~D.} \bibnamefont{Bandrauk}},
  \bibnamefont{and} \bibinfo{author}{\bibfnamefont{P.~B.}
  \bibnamefont{Corkum}}, \emph{\bibinfo{title}{Photon-momentum transfer in
  multiphoton ionization and in time-resolved holography with photoelectrons}},
  \bibinfo{journal}{Phys.~Rev.~A} \textbf{\bibinfo{volume}{92}},
  \bibinfo{pages}{051401(R)} (\bibinfo{year}{2015}).

\bibitem[{\citenamefont{Ivanov}(2014)}]{cuspm}
\bibinfo{author}{\bibfnamefont{I.~A.} \bibnamefont{Ivanov}},
  \emph{\bibinfo{title}{Evolution of the transverse photoelectron-momentum
  distribution for atomic ionization driven by a laser pulse with varying
  ellipticity}}, \bibinfo{journal}{Phys. Rev. A} \textbf{\bibinfo{volume}{90}},
  \bibinfo{pages}{013418} (\bibinfo{year}{2014}).

\bibitem[{\citenamefont{Ivanov and Kheifets}(2013)}]{circ6}
\bibinfo{author}{\bibfnamefont{I.~A.} \bibnamefont{Ivanov}} \bibnamefont{and}
  \bibinfo{author}{\bibfnamefont{A.~S.} \bibnamefont{Kheifets}},
  \emph{\bibinfo{title}{Time delay in atomic photoionization with circularly
  polarized light}}, \bibinfo{journal}{Phys. Rev. A}
  \textbf{\bibinfo{volume}{87}}, \bibinfo{pages}{033407}
  (\bibinfo{year}{2013}).

\bibitem[{\citenamefont{Ivanov and Kheifets}(2014)}]{circ7}
\bibinfo{author}{\bibfnamefont{I.~A.} \bibnamefont{Ivanov}} \bibnamefont{and}
  \bibinfo{author}{\bibfnamefont{A.~S.} \bibnamefont{Kheifets}},
  \emph{\bibinfo{title}{Strong-field ionization of he by elliptically polarized
  light in attoclock configuration}}, \bibinfo{journal}{Phys. Rev. A}
  \textbf{\bibinfo{volume}{89}}, \bibinfo{pages}{021402}
  (\bibinfo{year}{2014}).

\bibitem[{\citenamefont{Landau and Lifshitz}(1977)}]{LL3}
\bibinfo{author}{\bibfnamefont{L.~D.} \bibnamefont{Landau}} \bibnamefont{and}
  \bibinfo{author}{\bibfnamefont{E.~M.} \bibnamefont{Lifshitz}},
  \emph{\bibinfo{title}{Quantum Mechanics}} (\bibinfo{publisher}{Pergamon
  Press, New York}, \bibinfo{year}{1977}).

\bibitem[{\citenamefont{Kheifets and Ivanov}(2010)}]{delay1}
\bibinfo{author}{\bibfnamefont{A.~S.} \bibnamefont{Kheifets}} \bibnamefont{and}
  \bibinfo{author}{\bibfnamefont{I.~A.} \bibnamefont{Ivanov}},
  \emph{\bibinfo{title}{Delay in atomic photoionization}},
  \bibinfo{journal}{Phys. Rev. Lett.}
  \textbf{\bibinfo{volume}{105}}(\bibinfo{number}{23}), \bibinfo{pages}{233002}
  (\bibinfo{year}{2010}).

\bibitem[{\citenamefont{Ivanov}(2012)}]{mined2}
\bibinfo{author}{\bibfnamefont{I.~A.} \bibnamefont{Ivanov}},
  \emph{\bibinfo{title}{Double photoionization of the hydrogen molecule from
  the viewpoint of the time-delay theory}}, \bibinfo{journal}{Phys. Rev. A}
  \textbf{\bibinfo{volume}{86}}, \bibinfo{pages}{023419}
  (\bibinfo{year}{2012}).
	
\bibitem[{\citenamefont{Goldberger and Watson}(1964)}]{gw64}
\bibinfo{author}{\bibfnamefont{M.~L.} \bibnamefont{Goldberger}}
  \bibnamefont{and} \bibinfo{author}{\bibfnamefont{K.~M.}
  \bibnamefont{Watson}}, \emph{\bibinfo{title}{Collision theory}}
  (\bibinfo{publisher}{John Wiley and Sons}, \bibinfo{address}{New York},
  \bibinfo{year}{1964}).
	
\bibitem[{\citenamefont{Wigner}(1955)}]{delay2}
\bibinfo{author}{\bibfnamefont{E.~P.} \bibnamefont{Wigner}},
  \emph{\bibinfo{title}{Lower limit for the energy derivative of the scattering
  phase shift}}, \bibinfo{journal}{Phys. Rev.} \textbf{\bibinfo{volume}{98}},
  \bibinfo{pages}{145} (\bibinfo{year}{1955}).
	
	
\bibitem[{\citenamefont{Serov et~al.}(2019)\citenamefont{Serov, Bray, and
  Kheifets}}]{PhysRevA.99.063428}
\bibinfo{author}{\bibfnamefont{V.~V.} \bibnamefont{Serov}},
  \bibinfo{author}{\bibfnamefont{A.~W.} \bibnamefont{Bray}}, \bibnamefont{and}
  \bibinfo{author}{\bibfnamefont{A.~S.} \bibnamefont{Kheifets}},
  \emph{\bibinfo{title}{Numerical attoclock on atomic and molecular hydrogen}},
  \bibinfo{journal}{Phys. Rev. A} \textbf{\bibinfo{volume}{99}},
  \bibinfo{pages}{063428} (\bibinfo{year}{2019}).

\end{thebibliography}

\end{document}